%% file: allerton2012.tex
\documentclass[conference]{IEEEtran}

\IEEEoverridecommandlockouts
\input{preamble}

\begin{document}
\pgfdeclarelayer{background}
\pgfdeclarelayer{foreground}
\pgfsetlayers{background,main,foreground}

\title{\vspace{0.25in} A Proof of Threshold Saturation for Spatially-Coupled LDPC Codes on BMS Channels}

\author{Santhosh Kumar$^{\dagger}$, Andrew J. Young$^{\dagger}$, Nicolas Macris$^{\ddagger}$, and Henry D. Pfister$^{\dagger}$
  \thanks{This material is based upon work supported in part by the
    National Science Foundation (NSF) under Grant No. 0747470.
    The work of N.~Macris was supported by Swiss National Foundation Grant No. 200020-140388.
    Any opinions, findings, conclusions, and recommendations expressed
    in this material are those of the authors and do not necessarily
    reflect the views of these sponsors.}
   \\Department of Electrical and Computer Engineering, Texas A\&M University$^{\dagger}$
   \\School of Computer and Communication Sciences, \'{E}cole Polytechnique F\'{e}d\'{e}rale de Lausanne$^{\ddagger}$
}

\maketitle

\begin{abstract}
  Low-density parity-check (LDPC) convolutional codes have been shown to exhibit excellent performance under low-complexity belief-propagation decoding~\cite{Sridharan-aller04,Lentmaier-isit05}.
  This phenomenon is now termed threshold saturation via spatial coupling.
  The underlying principle behind this appears to be very general and spatially-coupled (SC) codes have been successfully applied in numerous areas.
  Recently, SC regular LDPC codes have been proven to achieve capacity universally, over the class of binary memoryless symmetric (BMS) channels, under belief-propagation decoding~\cite{Kudekar-it11,Kudekar-arxiv12}.

  In \cite{Yedla-istc12,Yedla-itw12}, potential functions are used to prove that the BP threshold of SC irregular LDPC ensembles saturates, for the binary erasure channel, to the conjectured MAP threshold (known as the Maxwell threshold) of the underlying irregular ensembles.
  In this paper, that proof technique is generalized to BMS channels, thereby extending some results of~\cite{Kudekar-arxiv12} to irregular LDPC ensembles.
  We also believe that this approach can be expanded to cover a wide class of graphical models whose message-passing rules are associated with a Bethe free energy.
\end{abstract}


\input{allerton_intro}

\section{Preliminaries}
\input{preliminaries}

\section{Single System}
\input{single_system}

\section{Coupled System}
\input{coupled_system}

\section{A Proof of Threshold Saturation}
\input{maintheorem}

\section{Conclusions}

In this paper, the proof technique based on potential functions (in~\cite{Yedla-istc12,Yedla-itw12}) is extended to BMS channels.
This extends the results of~\cite{Kudekar-arxiv12} by proving threshold saturation to the Maxwell threshold for BMS channels and the class of SC irregular LDPC ensembles.
In particular, for any family of BMS channels $\msc(\h)$ that is ordered by degradation and parameterized by entropy, $\h$, a potential threshold $\h^*$ is defined that satisfies $\hmap \leq \h^*$, where $\hmap$ defines the MAP threshold of the underlying LDPC($\lambda,\rho$) ensemble.

The main result is that, for sufficiently large $w$, the SC DE equation converges to the perfect decoding fixed point for any BMS channel $\msc(\h)$ with $\h < \h^{*}$.
The approach taken in this paper can be seen as analyzing the average Bethe free energy of the SC ensemble in the large system limit~\cite{Macris-it07,Mori-isit11}.

This result reiterates the generality of the threshold saturation phenomenon, which is now evident from the many observations and proofs that span a wide variety of systems.
We also believe that this approach can be extended to more general graphical models by computing the average Bethe free energy of the corresponding SC system.

\subsubsection*{Acknowledgment}
The authors thank R{\"u}diger Urbanke, Arvind Yedla, and Yung-Yih Jian for a number of very useful discussions during the early stages of this research.
\appendix
\input{appendix}

\bibliographystyle{ieeetr}
\bibliography{WCLabrv,WCLbib,WCLnewbib}

\end{document}

%% file: preamble.tex
\usepackage[margin=0.75in]{geometry}
\usepackage{amsmath,amssymb,bbm,stmaryrd}
\usepackage{bm}
\usepackage{tikz}
\usepackage{enumerate}
\usepackage{graphicx}
\usepackage{stfloats}
\usepackage{mathtools}
\usepackage{float}
\usetikzlibrary{%
  arrows,%
  shapes.misc,
  shapes.arrows,%
  shapes,%
  chains,%
  matrix,%
  positioning,
  scopes,%
  decorations.pathmorphing,
  shadows,%
  backgrounds,%
  fit,%
  petri%
}
\usepackage{pgfplots}
\usepackage{flushend}
\usepackage{scalefnt}

\newcommand{\h}{\texttt{h}}

\newcommand{\hmap}{\h^{\mathrm{MAP}}}
\renewcommand{\epsilon}{\varepsilon}

\global\long\def\expt{\mathbb{E}}

\newcommand{\bop}{\ast}
\newcommand{\vnop}{\varoast}
\newcommand{\cnop}{\boxast}
\newcommand{\diff}[1]{d#1}

\newcommand{\mb}[1]{\mathbf{#1}}
\newcommand{\mbb}[1]{\mathbb{#1}}

\newcommand{\mc}[1]{\mathcal{#1}}
\newcommand{\ms}[1]{\mathsf{#1}}

\newcommand{\mse}{\mathsf{e}}
\newcommand{\msx}{\mathsf{x}}
\newcommand{\msxvn}{\tilde{\mathsf{x}}}
\newcommand{\msy}{\mathsf{y}}
\newcommand{\msz}{\mathsf{z}}

\newcommand{\msc}{\mathsf{c}}
\newcommand{\msbx}{\underline{\mathsf{x}}}
\newcommand{\msby}{\underline{\mathsf{y}}}
\newcommand{\msbxcn}{\tilde{\underline{\mathsf{x}}}}
\newcommand{\msbz}{\underline{\mathsf{z}}}

\newcommand{\extR}{\overline{\mathbb{R}}}

\newcommand{\vnunit}{\Delta_0}
\newcommand{\cnunit}{\Delta_\infty}

\newcommand{\deri}[1]{\mathrm{d}_{ #1 }\hspace{0.05cm}}
\newcommand{\dderi}[1]{\mathrm{d}_{ #1 }^2\hspace{0.05cm}}

\newcommand{\ent}[1]{ \mathrm{H} \left( #1 \right) }

\newcommand{\degr}{\succ}
\newcommand{\degreq}{\succeq}
\newcommand{\upgr}{\prec}
\newcommand{\upgreq}{\preceq}

\newcommand{\meass}{\mathcal{M}}
\newcommand{\probs}{\mathcal{X}}
\newcommand{\dpros}{\mathcal{X}_{\mathrm{d}}}

\newcommand{\chend}{N_{w}}

\newtheorem{theorem}{Theorem}
\newtheorem{lemma}[theorem]{Lemma}
\newtheorem{proposition}[theorem]{Proposition}
\newtheorem{definition}[theorem]{Definition}

\newtheorem{remark}[theorem]{Remark}
\newtheorem{corollary}[theorem]{Corollary}

\providecommand{\abs}[1]{\left\lvert#1\right\rvert}

\newlength\tikzwidth
\newlength\tikzheight



%% file: allerton_intro.tex
\section{Introduction}

Low-density parity-check (LDPC) convolutional codes were introduced in \cite{Felstrom-it99} and shown to have outstanding performance under belief-propagation (BP) decoding in \cite{Sridharan-aller04,Lentmaier-isit05,Lentmaier-it10}.
The fundamental principle behind this phenomenon is described by Kudekar, Richardson, and Urbanke in \cite{Kudekar-it11} and coined \emph{threshold saturation} via spatial coupling.
For the binary erasure channel (BEC), they prove that spatially coupling a collection of $(d_v,d_c)$-regular LDPC ensembles produces a new (nearly) $(d_v,d_c)$-regular ensemble whose BP threshold approaches the MAP threshold of the original ensemble.
Recently, a proof of threshold saturation (to the ``area threshold'') has been given for $(d_v,d_c)$-regular LDPC ensembles on binary memoryless symmetric (BMS) channels when $d_v/d_c$ is fixed and $d_v,d_c$ are sufficiently large~\cite{Kudekar-arxiv12}.
This result implies that SC-LDPC codes achieve capacity universally over the class of BMS channels because the ``area threshold'' of regular LDPC codes approaches the Shannon limit uniformly over this class when $d_v/d_c$ is fixed and $d_v,d_c$ are increased.

The idea of threshold saturation via spatial coupling has recently started a small revolution in coding theory, and SC codes have now been observed to \emph{universally} approach the capacity regions of many systems~\cite{Lentmaier-it10,Kudekar-istc10,Rathi-isit11,Yedla-isit11,Kudekar-isit11-DEC,Kudekar-isit11-MAC,Nguyen-arxiv11,Nguyen-icc12}.
For spatially-coupled systems with suboptimal component decoders, such as message-passing decoding of code-division multiple access (CDMA)~\cite{Takeuchi-isit11,Schlegel-isit11} or iterative hard-decision decoding of SC generalized LDPC codes~\cite{Jian-isit12}, the threshold saturates instead to an intrinsic threshold defined by the suboptimal component decoders.

SC has also led to new results for $K$-SAT, graph coloring, and the Curie-Weiss model in statistical physics~\cite{Hassani-itw10,Hassani-jsm12,Hassani-arxiv11}.
For compressive sensing, SC measurement matrices were introduced in~\cite{Kudekar-aller10}, shown to give large improvements with Gaussian approximated BP reconstruction in~\cite{Krzakala-arxiv11}, and finally proven to achieve the information-theoretic limit in~\cite{Donoho-arxiv11}.
Recent results based on spatial coupling are now too numerous to cite thoroughly.

A different proof technique, based on potential functions, is used in \cite{Yedla-istc12} to prove that the BP threshold of spatially-coupled (SC) irregular LDPC ensembles saturates to the conjectured MAP threshold (known as the Maxwell threshold) of the underlying irregular ensembles.
This technique is closely related to the analysis in \cite{Hassani-itw10,Hassani-jsm12} for the Curie-Weiss model, the heuristic approach in \cite{Takeuchi-ieice12}, and the continuum approach used to prove threshold saturation for compressed sensing in~\cite{Donoho-arxiv11}.
In this paper, the proof technique based on potential functions (in~\cite{Yedla-istc12,Yedla-itw12}) is extended to BMS channels.
This extends the results of~\cite{Kudekar-arxiv12} by proving threshold saturation to the Maxwell threshold for BMS channels and a wide class of SC irregular LDPC ensembles.
The main result is summarized in the following theorem whose proof comprises the majority of this paper.
\begin{theorem}
Consider the SC LDPC($\lambda,\rho$) ensemble defined in Section~\ref{sec:spatial_coupling} and a family of BMS channels $\msc(\h)$ that is ordered by degradation and parameterized by entropy, $\h$.
For any BMS channel $\msc(\h)$ with $\h$ less than the potential threshold $\h^*$, there exists a sufficiently large coupling parameter $w_{0}$ such that, for all $w > w_{0}$, the SC density evolution converges to the perfect decoding solution.
\end{theorem}

Many observations, formal proofs, and a large variety of applications systems bear evidence to the generality of threshold saturation.
In particular, the approach taken in this paper can be seen as analyzing the average Bethe free energy of the SC ensemble in the large system limit~\cite{Macris-it07,Mori-isit11}. 
Therefore, it is tempting to conjecture that this approach can be applied to more general graphical models by computing the average Bethe free energy of the corresponding SC system.


%% file: preliminaries.tex
\label{section:preliminaries}
\subsection{Measure Algebras}
\label{subsection:Operators}

We call a Borel measure $\msx$ symmetric if 
\begin{equation*}
  \int_{-E} e^{-\alpha / 2} \, \msx ( \diff{\alpha} )  = \int_{E} e^{ - \alpha / 2} \, \msx( \diff{\alpha} ) ,
\end{equation*}
for all Borel sets $E \subset \extR$ where one of the integrals is finite.
Let $\meass = M(\extR)$ be the space of finite signed symmetric Borel measures on the extended real numbers $\extR$.
In this work, the primary interest is on convex combinations and differences of symmetric probability measures that inherit many of their properties from $\meass$.
Let $\probs \subset \meass$ be the convex subset of symmetric probability measures.
Also, let $\dpros \subset \meass$ be the subset of differences of symmetric probability measures:
\begin{align*}
  \dpros \triangleq \left\{ \msx_1 - \msx_2 \mid \msx_1, \msx_2 \in \probs \right\} .
\end{align*}
In the interest of notational consistency, $ \msx $ is reserved for both finite signed symmetric Borel measures and symmetric probability measures, and $ \msy $, $ \msz $ denote differences of symmetric probability measures.

In this space, there are two important binary operators, $\vnop$ and $\cnop$, that denote the variable-node and check-node density evolution operations for log-likelihood ratio (LLR) message distributions, respectively.
The wildcard $\bop$ is used to represent either operator in statements that apply to both operations.
For example, the shorthand $\msx^{\bop n}$ is used to denote
\begin{equation*}
  \underbrace{\msx \bop \cdots \bop \msx}_{n}.
\end{equation*}
For a polynomial $\lambda(\alpha) = \sum_{n=0}^{\mathrm{deg}(\lambda)} \lambda_{n} \alpha^{n}$ with real coefficients, we define
\begin{equation*}
  \lambda^{\bop}(\msx) \triangleq \sum_{n=0}^{\mathrm{deg}(\lambda)} \lambda_{n} \msx^{\bop n}.
\end{equation*}

Now, we give an explicit integral characterization of the operators $\vnop$ and $\cnop$.
For $\msx_{1}, \msx_2 \in \meass$, and any Borel set $E \subset \extR$, define
\begin{align*}
  ( \msx_{1} \vnop \msx_{2} )(E) &\triangleq \int \msx_{1}( E-\alpha ) \, \msx_{2} ( \diff{\alpha} ), \\
  ( \msx_{1} \cnop \msx_{2} )(E) &\triangleq \int \msx_{1} \left( 2 \tanh^{-1} \left( \frac{\tanh(\frac{E}{2})}{\tanh(\frac{\alpha}{2})} \right) \right) \msx_{2} ( \diff{\alpha} ).
\end{align*}
Associativity, commutativity, and linearity follow. Therefore, for measures $\msx_{1}$, $\msx_{2}$,  $\msx_{3} \in \meass$ and scalars $\alpha_{1}$, $\alpha_{2} \in \mbb{R}$,
\begin{align*}
  \msx_{1} \bop \left( \msx_{2} \bop \msx_{3} \right) &= \left( \msx_{1} \bop \msx_{2} \right) \bop \msx_{3} , \\
  \msx_{1} \bop \msx_{2} &= \msx_{2} \bop \msx_{1} , \\ 
  (\alpha_{1} \msx_{1} + \alpha_{2} \msx_{2}) \bop \msx_{3} &= \alpha_{1} (\msx_{1} \bop \msx_{3}) + \alpha_{2} ( \msx_{2} \bop \msx_{3} ).
\end{align*}
Moreover, the space of symmetric probability measures is closed under these binary operations~\cite{RU-2008}.
In a more abstract sense, the measure space $\meass$ along with the multiplication operator $\bop$ forms a commutative unital associative algebra and this algebraic structure is induced on the space of symmetric probability measures.
There is also an intrinsic connection between the algebras defined by each operator and one consequence is the duality (or conservation) result in Proposition \ref{proposition:duality}.
For the multiplicative identities in these algebras, $\mse_{\vnop} = \vnunit$ and $\mse_{\cnop} = \cnunit$, we define $\msx^{\bop 0} = \mse_\bop$ and observe the following relationships under the dual operation:
\begin{align*}
  \vnunit \cnop \ms{x} &= \vnunit
  &
  \cnunit \vnop \ms{x} &= \cnunit.
\end{align*}
In general, however, these operators do not associate
\begin{align*}
  \msx_{1} \vnop (\msx_{2} \cnop \msx_{3}) &\neq (\msx_{1} \vnop \msx_{2}) \cnop \msx_{3} \\
  \msx_{1} \cnop (\msx_{2} \vnop \msx_{3}) &\neq (\msx_{1} \cnop \msx_{2}) \vnop \msx_{3},
\end{align*}
nor distribute
\begin{align*}
  \msx_{1} \vnop (\msx_{2} \cnop \msx_{3}) &\neq (\msx_{1} \vnop \msx_{2}) \cnop ( \msx_{1} \vnop \msx_{3} )\\
  \msx_{1} \cnop (\msx_{2} \vnop \msx_{3}) &\neq (\msx_{1} \cnop \msx_{2}) \vnop ( \msx_{1} \cnop \msx_{3} ).
\end{align*}

\subsection{Partial Ordering by Degradation}
\label{subsection:Degradation}
Degradation is an important concept that allows one to compare some LLR message distributions.
The order imposed by degradation is indicative of relating probability measures through a communication channel \cite[Definition 4.69]{RU-2008}.
\begin{definition}
  \label{definition:degradation}
  For $\msx \in \probs$ and $f:[0,1]\to\mathbb{R}$, define
  \begin{align*}
    I_{f}(\msx)  \triangleq \int f \left( \abs{ \tanh\left(\tfrac{\alpha}{2}\right) } \right) \msx( \diff{\alpha} ) .
  \end{align*}
  For $\msx_{1}, \msx_{2} \in \probs$, $\msx_{1}$ is said to be \emph{degraded} with respect to $\msx_{2}$ (denoted $\msx_{1} \degreq \msx_{2}$), if $I_{f}(\msx_1) \geq I_{f}(\msx_2)$ for all concave non-increasing $f$.
  Furthermore, $\msx_{1}$ is said to be \emph{strictly degraded} with respect to $\msx_{2}$ (denoted $\msx_{1} \degr \msx_{2}$) if $\msx_{1} \degreq \msx_{2}$ and there is a concave non-increasing $f$ so that $I_{f}(\msx_1) > I_{f}(\msx_2)$.
\end{definition}

Degradation defines a partial order, on the space of symmetric probability measures, with maximal element $\vnunit$ and minimal element $\cnunit$.
In this paper, the notation for real intervals is overloaded and, for example, a half-open interval of measures is denoted by
\begin{equation*}
  (\msx_1,\msx_2] \triangleq \{ \msx' \in \probs \mid \msx_1 \upgr \msx' \upgreq \msx_2 \}.
\end{equation*}
This partial ordering is also preserved under the binary operations as follows.
\begin{proposition}[{\cite[Lemma 4.80]{RU-2008}}]
  \label{proposition:degradation_preservation}
  Let $\msx_{1}, \msx_{2} \in \probs$ be ordered by degradation $\msx_{1} \degreq \msx_{2}$.
  For any $\msx_{3} \in \mc{X}$, one has
  \begin{equation*}
    \msx_{1} \bop \msx_{3} \degreq \msx_{2} \bop \msx_{3}.
  \end{equation*}
\end{proposition}
This ordering is our primary tool in describing relative channel quality, and thresholds.
For further information see  \cite[pp.~204-208]{RU-2008}.

\subsection{Entropy Functional for Symmetric Measures}
Entropy is a fundamental quantity in information theory and communication, and it is defined as the linear functional, $\mathrm{H}: \meass \rightarrow \mbb{R}$, given by the integral
\begin{align*}
  \ent{\msx} \triangleq \int \log_2 \left(1 + e^{-\alpha} \right) \msx(\diff{\alpha}).
\end{align*}
The entropy functional is the primary functional used in our analysis.
It exhibits an order under degradation and, for symmetric probability measures $\msx_{1} \degr \msx_{2}$, one has
\begin{align*}
  \ent{ \msx_{1} }  > \ent{ \msx_{2} }.
\end{align*}
The restriction to symmetric probability measures ($\msx \in \probs$) also implies the bound
\begin{equation*}
  0 \leq \ent{\msx} \leq 1 .
\end{equation*}
The operators $\vnop$ and $\cnop$ admit a number of relationships under the entropy functional. 
The following results will prove invaluable in the ensuing analysis.
Proposition \ref{proposition:duality} provides an important conservation result (also known as the duality rule for entropy) and Proposition \ref{proposition:duality_difference} extends this relation to encompass differences of symmetric probability measures.
\begin{proposition}[{\cite[pp.~196]{RU-2008}}]
  \label{proposition:duality}
  For $\msx_{1},\msx_{2} \in \probs$, one has
  \begin{equation}
    \ent{ \msx_{1} \vnop \msx_{2} } + \ent{ \msx_{1} \cnop \msx_{2} } = \ent{ \msx_{1} } + \ent{ \msx_{2} }.
  \end{equation}
\end{proposition}

\begin{proposition}
  \label{proposition:duality_difference}
  For $\msx_{1}$, $\msx_{2}$, $\msx_{3}$, $\msx_{4} \in \probs$, one finds that
  \begin{equation*}
    \quad \ent{ \msx_{1} \vnop (\msx_{3} - \msx_{4}) }  + \ent{ \msx_{1} \cnop (\msx_{3} - \msx_{4}) } 
    = \ent{ \msx_{3} - \msx_{4} },
  \end{equation*}
  \begin{equation*}
    \ent{  (\msx_{1} - \msx_{2}) \vnop (\msx_{3} - \msx_{4}) }  + \ent{ (\msx_{1} - \msx_{2}) \cnop (\msx_{3} - \msx_{4}) } = 0.
  \end{equation*}
\end{proposition}

\begin{IEEEproof}
  Consider the LHS of the first equality,
  \begin{align*}
    & \ent{ \msx_{1} \vnop (\msx_{3} - \msx_{4}) } + \ent{ \msx_{1} \cnop (\msx_{3} - \msx_{4}) } \\
    &=  \ent{ \msx_{1} \vnop \msx_{3} } + \ent{ \msx_{1} \cnop \msx_{3} } - \ent{ \msx_{1} \vnop \msx_{4} } - \ent{ \msx_{1} \cnop \msx_{4} } \\
    &= \ent{ \msx_{1} } + \ent{ \msx_{3} } - \ent{ \msx_{1} } - \ent{ \msx_{4} }  \quad \text{(Proposition \ref{proposition:duality})}  \\
    &= \ent{ \msx_{3}- \msx_{4} }.
  \end{align*}
  The second equality follows by expanding the LHS and applying the first equality twice.
\end{IEEEproof}

Due to the symmetry of the measures the entropy functional has an equivalent series representation.
\begin{proposition}
  \label{proposition:entropy_symmetric_measures}
  If $\msx \in \meass$, then 
  \begin{align*}
    \ent{\msx} = \msx \left( \extR \right) - \sum_{k=1}^{\infty} \frac{\left( \log 2 \right)^{-1}}{2k (2k-1)} \int \left( \tanh \frac{\alpha}{2} \right)^{2k} \msx(\diff{\alpha}) .
  \end{align*}
\end{proposition}
\begin{IEEEproof}
  For a sketch of the proof, see \cite[Lemma 3]{Montanari-it05}.
\end{IEEEproof}
Define the linear functional
\begin{align*}
  M_{k}(\msx) \triangleq \int \left( \tanh \frac{\alpha}{2} \right)^{2k} \msx(\diff{\alpha}) .
\end{align*}
Since $-x^{2k}$ is a concave decreasing function over the interval $[0,1]$, for symmetric probability measures $\msx_1 \degreq \msx_2$, by Definition \ref{definition:degradation},
\begin{align*}
  M_{k}(\msx_1) \leq M_{k}(\msx_2) .
\end{align*}
Moreover, for $\msx \in \probs$,
\begin{align*}
  0 \leq M_{k}(\msx) \leq 1 .
\end{align*}
It is also easy to see that the functional $M_{k}(\cdot)$ takes the following product form under the operator $\cnop$,
\begin{align*}
  M_{k}(\msx_1 \cnop \msx_2) = M_{k}(\msx_1) M_{k}(\msx_2) .
\end{align*}
Also, if $\msy_1, \msy_2 \in \dpros$ are the differences of symmetric probability measures, then $\msy_{1}(\extR) = \msy_{2}(\extR) = (\msy_1 \bop \msy_2) (\extR) = 0$. 
This leads to the following result.
\begin{proposition}
  \label{proposition:entropy_simplification}
  For $\msy_1, \msy_2 \in \dpros$, the entropy functional reduces to 
  \begin{align*}
    \ent{\msy_{1}} &= - \sum_{k=1}^{\infty} \frac{\left( \log 2 \right)^{-1}}{2k (2k-1)} M_{k}(\msy_{1}) , \\
    \ent{\msy_1 \cnop \msy_2} &= - \sum_{k=1}^{\infty} \frac{\left( \log 2 \right)^{-1}}{2k (2k-1)} M_{k}(\msy_1) M_{k}(\msy_2) .
  \end{align*}
  In particular, 
\begin{align*}
    \ent{\msy_{1} \cnop \msy_{1}} &= - \sum_{k=1}^{\infty} \frac{\left( \log 2 \right)^{-1}}{2k (2k-1)} M_{k}(\msy_{1})^{2} \leq 0 .
  \end{align*}
  with equality iff $\msy_1=0$.
\end{proposition}
The above proposition implies an important upper bound on the entropy functional for differences of symmetric probability measures under \emph{either} operator, $\vnop$ or $\cnop$.
\begin{proposition}
  \label{proposition:entropy_bound}
  If $\msx_1, \msx_1', \msx_2, \msx_3 \in \probs$ with $\msx_1' \degreq \msx_1$, then
  \begin{align*}
    \abs{\ent{ \left( \msx_1' - \msx_1 \right) \bop \left( \msx_2 - \msx_3 \right) }} \leq \ent{ \msx_1' - \msx_1 } .
  \end{align*}
\end{proposition}
\begin{IEEEproof}
  We show the result for the operator $\cnop$. The extension to $\vnop$ follows from the duality rule for entropy for differences of symmetric probability measures, Proposition \ref{proposition:duality_difference}.
  From Proposition \ref{proposition:entropy_simplification}
\begin{align*}
  & \abs{\ent{\left( \msx_1' - \msx_1 \right)  \cnop \left( \msx_2 - \msx_3 \right) }} \\
  & \quad \le \sum_{k=1}^{\infty} \frac{\left( \log 2 \right)^{-1}}{2k (2k-1)} \abs{ M_{k}(\msx_1' - \msx_1)}\abs{ M_{k}(\msx_2 - \msx_3)} \\
  & \quad \overset{(a)}{=} - \sum_{k=1}^{\infty} \frac{\left( \log 2 \right)^{-1}}{2k (2k-1)} M_{k}(\msx_1' - \msx_1) \abs{ M_{k}(\msx_2 - \msx_3) }, \\
  & \quad \overset{(b)}{\leq} - \sum_{k=1}^{\infty} \frac{\left( \log 2 \right)^{-1}}{2k (2k-1)} M_{k}(\msx_1' - \msx_1) \\
  & \quad = \ent{\msx_1' - \msx_1} ,
\end{align*}
where $(a)$ follows from $M_{k}(\msx_1') \leq M_{k}(\msx_1)$ and $(b)$ follows since $0 \leq M_{k}(\msx_2), M_{k}(\msx_3) \leq 1$.
\end{IEEEproof}
\begin{corollary}
  \label{corollary:entropy_bound_3variables}
  For $\msx_1, \msx_1', \msx_2, \msx_3, \msx_4 \in \probs$ with $\msx_1' \degreq \msx_1$, one has
  \begin{align*}
    \abs{\ent{\left( \msx_1' - \msx_1 \right) \bop \left( \msx_2 - \msx_3 \right) \bop \msx_4}} \leq \ent{\msx_1' - \msx_1} .
  \end{align*}
\end{corollary}
\begin{IEEEproof}
  This follows from Proposition \ref{proposition:entropy_bound} by replacing $\msx_2$, $\msx_3$ with $\msx_2 \bop \msx_4$ and $\msx_3 \bop \msx_4$, respectively.
\end{IEEEproof}

\subsection{Directional Derivatives}
\label{subsection:DirectionalDerivative}
The main result in this paper is derived using potential theory and differential relations.
One can avoid the technical challenges of differentiation in the abstract space of measures by focusing on directional derivatives of functionals that map measures to real numbers.
\begin{definition}
  \label{definition:directional_derivative}
  Let $F:\meass \rightarrow \mbb{R}$ be a functional on $\mc{M}$.
  The \emph{directional derivative} of $F$ at $\msx$ in the direction $\msy$ is
  \begin{align*}
    \deri{\msx} F(\msx)[\msy] \triangleq \lim_{\delta \rightarrow 0} \frac{F(\msx + \delta \msy ) - F(\msx)}{\delta} ,
  \end{align*}
  whenever the limit exists.  
\end{definition}

This definition is naturally extended to higher-order directional derivatives using
\begin{align*}
  \mathrm{d}_{\ms{x}}^{n} F(\msx) [ \msy_{1}, \ldots, \msy_{n} ] \triangleq \deri{\msx} \left( \cdots \deri{\msx} \left( \deri{\msx}  F \left( \msx \right) [ \msy_{1} ] \right) [\msy_{2}]  \cdots \right) [\msy_{n}],
\end{align*}
and vectors of measures using, for $\msbx = [\msx_1,\cdots,\msx_m]$, 
\begin{align*}
  \deri{\msbx} F(\msbx)[\msby] &\triangleq \lim_{\delta \rightarrow 0} \frac{F(\msbx + \delta \msby ) - F(\msbx)}{\delta},
\end{align*}
whenever the limit exists.
Similarly, we can define higher-order directional derivatives of functionals on vectors of measures.

The utility of directional derivatives for linear functionals is evident from the following Lemma.
\begin{lemma}
  \label{lemma:Example}
  Let $F : \meass \rightarrow \mbb{R}$ be a linear functional and $\bop$ be an associative, commutative, and linear binary operator.
  Then, for $\msx,\msy, \msz \in \meass$, we have
  \begin{align*}
    \deri{\msx} F(\msx^{\bop n})[\msy] &= n F( \msx^{\bop (n-1)} \bop \msy ) , \\
    \dderi{\msx} F(\msx^{\bop n})[\msy,\msz] &= n \left( n-1 \right) F \left( \msx^{\bop (n-2)} \bop \msy \bop \msz  \right) .
  \end{align*}
\end{lemma}
\begin{IEEEproof}
  A binary operation $\bop$ that is associative, commutative, and linear admits a binomial expansion of the form
  \begin{align*}
    \left( \msx + \delta \msy \right)^{\bop n} = \sum_{i=0}^{n} \delta^{i} \binom{n}{i} \msx^{\bop (n-i)} \bop \msy^{\bop i}.
  \end{align*}
  Then, the linearity of $F$ implies that
  \begin{align*}
    & F \left( (\msx + \delta \msy)^{\bop n} \right) - F \left( \msx^{\bop n} \right)  \\
    & \quad = \delta n F( \msx^{\bop (n-1)} \bop \msy ) + \sum_{i=2}^{n} \delta^{i} F( \msx^{\bop (n-i)} \bop \msy^{\bop i} ).
  \end{align*}
  Dividing by $\delta$ and taking the limit gives
  \begin{align*}
    \deri{\msx} F( \msx^{\bop n} )  [\msy] = n F( \msx^{\bop (n-1)} \bop \msy ).
  \end{align*}
  An analogous argument shows the result for the second-order directional derivative.
\end{IEEEproof}

\begin{remark}
  In general, applying Taylor's theorem to some mapping $T: \probs \to \probs$ requires advanced mathematical machinery.
  However, in our problem, the entropy functional and its interplay with the operators $\vnop$ and $\cnop$ impose a polynomial structure on the functions of interest, obviating the need for Fr\'{e}chet derivatives.
  Therefore, Taylor's theorem becomes quite simple for parameterized linear functionals $\phi : [0,1] \rightarrow \mathbb{R}$ of the form
  \begin{align*}
    \phi(t) = F\left(\msx_1 + t (\msx_2 -\msx_1)\right).
  \end{align*}
\end{remark}


%% file: single_system.tex

\label{section:singlesystem}

Let LDPC($\lambda,\rho$) denote the LDPC ensemble with variable-node degree distribution $\lambda$ and check-node degree distribution $\rho$.
The edge-perspective degree distributions $\lambda,\rho$ have an equivalent representation in terms of the node-perspective degree distributions $L$, $R$, namely, 
\begin{align*}
  \lambda(\alpha) &= \frac{L'(\alpha)}{L'(1)}, & \rho(\alpha) &= \frac{R'(\alpha)}{R'(1)}.
\end{align*}
This differential relationship is crucial in developing an appropriate potential functional for the LDPC$(\lambda, \rho)$ ensemble.

\emph{Density evolution} (DE) characterizes the asymptotic performance of the LDPC$(\lambda,\rho)$ ensemble by describing the evolution of message distributions with iteration.
For this ensemble, the DE update is compactly described by
\begin{align*}
  \msxvn^{(\ell+1)} = \msc \vnop \lambda^{\vnop}(\rho^{\cnop}(\msxvn^{(\ell)})) ,
\end{align*}
where $\msxvn^{(\ell)}$ is the variable-node output distribution after $\ell$ iterations of message passing~\cite{Richardson-it01,RU-2008}.

We now develop the necessary definitions for the single-system potential framework.
Included are the potential functional,  fixed points, stationary points, the directional derivative  of the potential functional, and thresholds.

\begin{definition}
  Consider a family of BMS channels whose LLR distributions $\msc(\h): [0,1] \to \probs$ are ordered by degradation and parameterized by their entropy $\ent{\msc(\h)}=\h$.
  The \emph{BP threshold channel} of such a family is defined to be $\msc^{\mathrm{BP}} \triangleq \msc(\h^{\mathrm{BP}})$, where $\h^{\mathrm{BP}}  \triangleq$
  \begin{align*} 
    \sup \left\{ \h \!\in\! [0,1] |  \text{$\msx \in \probs$ and $\msx \!=\! \msc(\h)\!\vnop\! \lambda^{\vnop}(\rho^{\cnop}(\msx)) \!\Rightarrow\! \msx\!=\!\cnunit$}   \right\} . 
  \end{align*}
\end{definition}

\begin{definition}
  For a family of BMS channels, $\msc(\h)$, the MAP threshold is given by $\h^{\mathrm{MAP}} \triangleq$
  \begin{align*} 
    \inf \left\{ \h \in [0,1] \mid \liminf_{n\to\infty} \tfrac{1}{n} \expt \left[ \ent{ X^n \mid Y^n (\msc(\h)) } \right] > 0 \right\} .
  \end{align*}
\end{definition}

\begin{definition}
  \label{def:uncoupled_potential}
  The \emph{potential functional} (or the average Bethe free energy), $U: \mc{X} \times \mc{X} \to \mbb{R}$, of the LDPC$(\lambda,\rho)$ ensemble is
  \begin{align*}
    U(\msx; \msc) &\triangleq  \tfrac{L'(1)}{R'(1)} \ent{R^{\cnop}(\msx)} + L'(1) \ent{ \rho^{\cnop}(\msx) } \\
    & \quad - L'(1) \ent{\msx \cnop \rho^{\cnop}(\msx)} - \ent{ \msc \vnop L^{\vnop} \left( \rho^{\cnop}(\msx) \right) } .
  \end{align*}
\end{definition}

\begin{figure}[t]
  \centering
  \setlength\tikzheight{5cm}
  \setlength\tikzwidth{6.5cm} 
  \input{./Figures/PotentialFunctionalBSC}
  \vspace{-3mm}
  \caption{ \label{figure:potential_functional} Potential functional for $(\lambda,\rho) = (x^2,x^5)$ on the binary symmetric channel (BSC), with $\texttt{h} \in \{0.40, 0.416, 0.44, 0.469, 0.48\}$. The $\msx$-input is chosen to be the binary AWGN channel (BAWGNC) with entropy $\h '$.}
\end{figure}
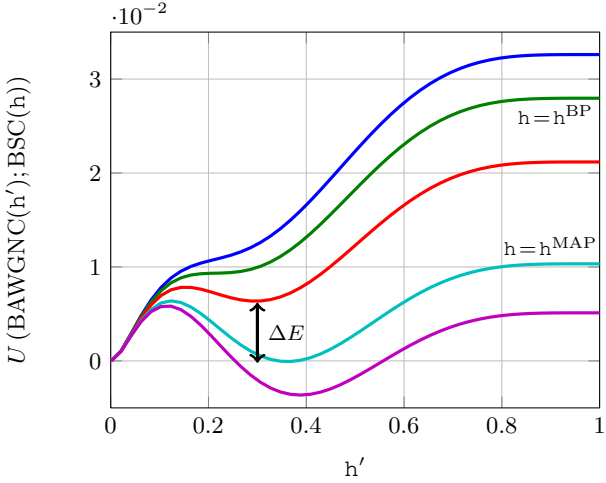

\begin{remark}
  This potential functional is essentially the negative of the replica-symmetric free energies calculated in~\cite{Montanari-it05,Macris-it07,Kudekar-it09}.
  When applied to the binary erasure channel, it is a constant multiple of the potential function defined in \cite{Yedla-istc12}.
  An example of $ U(\msx; \msc)$ is shown in Fig.~\ref{figure:potential_functional}.
\end{remark}

\begin{definition}
  \label{def:trial_entropy}
  The \emph{fixed-point potential}, $P: \mc{X} \to \mbb{R}$, of the LDPC$(\lambda,\rho)$ ensemble is
  \begin{align*}
    P(\msx) \triangleq U(\msx ; \ms{f}(\msx) ),
  \end{align*}
  where $\ms{f}: \mc{X} \to \mc{X}$ satisfies $\ms{f}(\msx) \vnop \lambda^{\vnop} \left( \rho^{\cnop}(\msx) \right) = \msx$.
\end{definition}

\begin{remark}
  For regular codes, one can use duality rule for entropy to show that the fixed-point potential defined above is exactly equal to the negative of the GEXIT integral functional denoted by $A$ in~\cite[Lemma 26]{Kudekar-arxiv12}.
  A similar relationship with EBP GEXIT integrals is believed to be true in general, but the authors are not aware of a proof for this.
\end{remark}

\begin{definition}
  For $\msx \in \probs$,
  \begin{enumerate}[a)]
  \item 
    $\msx$ is a \emph{fixed point} of \emph{density evolution} if 
    \begin{align*}
      \msx = \msc \vnop \lambda^{\vnop} \left( \rho^{\cnop}(\msx)  \right),
    \end{align*}
  \item
    $\msx$ is a \emph{stationary point} of the potential if, for all $\msy \in \dpros$,
    \begin{align*}
      \deri{\msx} U(\msx ; \msc) [\msy] = 0.
    \end{align*}
  \end{enumerate}
\end{definition}

\begin{lemma}
  For $\msx, \msc \in \probs$ and $\msy \in \dpros$, the directional derivative of the potential functional with respect to $\msx$ in the direction $\msy$ is
  \begin{align*}
    & \deri{\msx} U( \msx ; \msc ) [\msy] \\ 
    & \quad =  L'(1) \ent{ \left[ \msx - \msc \vnop \lambda^{\vnop} \left( \rho^{\cnop}(\msx) \right) \right] \vnop \left[ \rho'^{\cnop}(\msx) \cnop \msy \right] } .
  \end{align*}
\end{lemma}
\begin{IEEEproof}
  The directional derivative for each of the four terms is calculated following the procedure outlined in the proof of Lemma \ref{lemma:Example}.
  The first three terms are
  \begin{align*}
    \deri{\msx} \ent{ R^{\cnop}(\msx)} [\msy] &= R'(1) \ent{ \rho^{\cnop}(\msx) \cnop \msy }, \\
    \deri{\msx} \ent{ \rho^{\cnop}(\msx) }[\msy] &= \ent{ \rho'^{\cnop}(\msx) \cnop \msy }, \\    
    \deri{\msx}\ent{ \msx \cnop \rho^{\cnop}(\msx) } [\msy] &=  \ent{ \rho^{\cnop}(\msx) \cnop \msy }  + \ent{ \msx \cnop \rho'^{\cnop}(\msx) \cnop \msy } \\
    & \overset{(a)}{=}   \ent{ \rho^{\cnop}(\msx) \cnop \msy } + \ent{  \rho'^{\cnop}(\msx) \cnop \msy } \\
    & \qquad - \ent{ \msx \vnop \left[ \rho'^{\cnop}(\msx) \cnop \msy \right]},
  \end{align*}
  where $(a)$ follows from Proposition \ref{proposition:duality_difference}, $\rho'^{\cnop}(\msx) \cnop \msy$ is the difference of probability measures multiplied by the scalar $\rho'(1)$.
  Since the operators $\vnop$ and $\cnop$ do not associate, one must exercise care in analyzing the last term,
  \begin{align*}
    & \deri{\msx} \ent{ \msc \vnop L^{\vnop} \left( \rho^{\cnop}(\msx) \right) } [\msy] \\
    & \quad = L'(1) \ent{  \left[ \msc \vnop \lambda^{\vnop}(\rho^{\cnop}(\msx)) \right] \vnop \left[ \rho'^{\cnop}(\msx) \cnop \msy \right] }.
  \end{align*}
  Consolidating the four terms gives the desired result.
\end{IEEEproof}

\begin{corollary}
  \label{corollary:fixed_stationary}
  If $\msx \in \probs$ is a fixed point of density evolution, then it is also a stationary point of the potential functional.
\end{corollary}

\begin{definition}
  \label{definition:basin_of_attraction}
  For a channel $\msc \in \probs$, define the \emph{basin of attraction} for $\cnunit$ as
\begin{align*}
\mc{V}(\msc) \triangleq \left\{ \msx \in \probs \mid  \msx' \degr \msc \vnop \lambda^{\vnop} ( \rho^{\cnop} (\msx') ) \; \forall \, \msx' \in (\cnunit,\msx] \right\}.
\end{align*}
\end{definition}


\begin{definition}
  \label{definition:energy_gap}
  For a channel $\msc \in \probs$, the \emph{energy gap} is defined as
  \begin{align*}
    \Delta E (\msc) \triangleq \inf_{\msx \in \probs \setminus \mc{V}(\msc) } U(\msx;\msc).
  \end{align*}
\end{definition}

\begin{lemma}
  \label{lemma:monotonicity}
  The following relations are important in characterizing the thresholds.
  Suppose $\h_1 > \h_2$.
  Then
  \begin{enumerate}[a)]
    \item $\msc(\h_1) \degr \msc(\h_2)$
    \item $U(\msx ; \msc(\h_1)) < U(\msx ; \msc(\h_2))$, if $\msx \neq \cnunit$.
    \item $\mc{V}(\msc(\h_1)) \subseteq \mc{V}(\msc(\h_2)) \Longrightarrow \probs\setminus\mc{V}(\msc(\h_1)) \supseteq \probs\setminus\mc{V}(\msc(\h_2)) $
    \item $\Delta E(\msc(\h_1)) < \Delta E(\msc(\h_2))$
  \end{enumerate}
\end{lemma}
Lemma \ref{lemma:monotonicity}d holds under a minor restriction.
In particular, when there are no degree-two variables nodes.
See \cite{Kumar-itsub13} for details.

\begin{definition}
  \label{definition:potential_threshold_channel}
  The \emph{potential threshold channel} of a family, $\msc(\h)$, of BMS channels is given by $\msc^* \triangleq \msc(\h^*)$, where
  \begin{align*}
    \h^*  &\triangleq \sup \{ \h \in [\h^{\mathrm{BP}},1] \mid \Delta E (\msc(\h)) > 0 \}
  \end{align*}
\end{definition}
Note that, if $\h < \h^*$, from Lemma \ref{lemma:monotonicity}d, $\Delta E(\msc(\h)) > 0$.

\begin{lemma}
  \label{lemma:MAP_threshold}
  From~\cite{Montanari-it05,Macris-it07,Kudekar-it09}, the following holds for degree distributions without odd-degree checks on any BMS channel and any degree distribution on the binary-input AWGN channel:
  \begin{enumerate}[a)]
  \item $\displaystyle{\liminf_{n \to \infty}} \tfrac{1}{n} \expt \left[ \ent{ X^n | Y^n (\msc(\h)) } \right] \geq -\inf_{\msx \in \probs} U(\msx;\msc(\h)) ,$
  \item $\h^{\mathrm{MAP}} \leq \h^* .$
  \end{enumerate}
\end{lemma}
\begin{IEEEproof}
  Since the potential functional is the negative of the replica-symmetric free energies in~\cite{Montanari-it05,Macris-it07,Kudekar-it09}, the main result of these papers translates directly into the first result.
  For the second result, consider any $\h>\h^*$. 
  Below, we will arrive at a conclusion that $\h \geq \h^{\mathrm{MAP}}$, which establishes the desired result.
  Note that from Lemma \ref{lemma:monotonicity}d, $\Delta E(\msc(\h)) < 0$.
  From the first part of this lemma, 
  \allowdisplaybreaks{
    \begin{align*}
      & \displaystyle{\liminf_{n \to \infty}} \tfrac{1}{n} \expt \left[ \ent{ X^n | Y^n (\msc(\h)) } \right] \\
      & \quad \geq -\inf_{\msx \in \probs} U(\msx;\msc(\h)) \geq -\inf_{\msx \in \probs \setminus \mc{V}(\msc)} U(\msx;\msc(\h)) \\
      & \quad = - \Delta (E(\msc(\h))) > 0 .
    \end{align*}
  }
  Hence, $\h \geq \h^{\mathrm{MAP}}$.
\end{IEEEproof}

\begin{remark}
  This implies that the MAP threshold is upper bounded by the potential threshold for the single system.
  It is conjectured that, in general, these two quantities are actually equal.
  Some progress has been made towards proving this for special cases~\cite{Giurgiu-isit12}.
\end{remark}




%% file: Figures/PotentialFunctionalBSC.tex
%
%
\begin{tikzpicture}

\tikzstyle{every node}=[font=\small]

\definecolor{mycolor1}{rgb}{0,0.5,0}
\definecolor{mycolor2}{rgb}{0,0.75,0.75}
\definecolor{mycolor3}{rgb}{0.75,0,0.75}

\begin{axis}[%
scale only axis,
width=\tikzwidth,
height=\tikzheight,
xmin=0, xmax=1,
ymin=-0.005, ymax=0.035,
xlabel={$\h '$},
ylabel={$U\left( \mathrm{BAWGNC}(\h') ; \mathrm{BSC}(\h) \right)$},
xtick={0,0.2,0.4,0.6,0.8,1},
xmajorgrids,
ymajorgrids,
zmajorgrids,
]

\addplot [
color=blue,
solid,
line width=1.2pt,
]
coordinates{
 (0.001,4.13365e-06)(0.0213878,0.00107232)(0.0417755,0.00291826)(0.0621633,0.00481626)(0.082551,0.0064815)(0.102939,0.00781906)(0.123327,0.0088272)(0.143714,0.00955118)(0.164102,0.0100569)(0.18449,0.0104156)(0.204878,0.0106955)(0.225265,0.0109564)(0.245653,0.0112473)(0.266041,0.011606)(0.286429,0.0120589)(0.306816,0.0126216)(0.327204,0.0133008)(0.347592,0.014095)(0.36798,0.0149964)(0.388367,0.0159923)(0.408755,0.0170662)(0.429143,0.0181995)(0.449531,0.0193723)(0.469918,0.0205646)(0.490306,0.0217566)(0.510694,0.0229303)(0.531082,0.0240688)(0.551469,0.0251578)(0.571857,0.026185)(0.592245,0.0271406)(0.612633,0.0280172)(0.63302,0.02881)(0.653408,0.0295164)(0.673796,0.030136)(0.694184,0.0306704)(0.714571,0.0311229)(0.734959,0.0314984)(0.755347,0.031803)(0.775735,0.0320436)(0.796122,0.032228)(0.81651,0.0323644)(0.836898,0.0324609)(0.857286,0.0325256)(0.877673,0.0325661)(0.898061,0.0325892)(0.918449,0.0326007)(0.938837,0.0326053)(0.959224,0.0326064)(0.979612,0.0326065)(1,0.0326065) 
};

\addplot [
color=mycolor1,
solid,
line width=1.2pt,
]
coordinates{
 (0.001,4.13208e-06)(0.0213878,0.00106479)(0.0417755,0.00287635)(0.0621633,0.00470579)(0.082551,0.00626758)(0.102939,0.00746905)(0.123327,0.00831233)(0.143714,0.00884725)(0.164102,0.00914456)(0.18449,0.00928045)(0.204878,0.00932761)(0.225265,0.00935015)(0.245653,0.00940115)(0.266041,0.00952172)(0.286429,0.00974125)(0.306816,0.010078)(0.327204,0.0105408)(0.347592,0.0111296)(0.36798,0.011838)(0.388367,0.0126542)(0.408755,0.013562)(0.429143,0.0145433)(0.449531,0.015578)(0.469918,0.0166457)(0.490306,0.0177265)(0.510694,0.0188015)(0.531082,0.0198534)(0.551469,0.0208668)(0.571857,0.0218286)(0.592245,0.0227282)(0.612633,0.0235575)(0.63302,0.0243105)(0.653408,0.0249839)(0.673796,0.0255765)(0.694184,0.0260891)(0.714571,0.0265242)(0.734959,0.0268862)(0.755347,0.0271803)(0.775735,0.0274132)(0.796122,0.027592)(0.81651,0.0277245)(0.836898,0.0278183)(0.857286,0.0278814)(0.877673,0.0279208)(0.898061,0.0279433)(0.918449,0.0279546)(0.938837,0.0279591)(0.959224,0.0279602)(0.979612,0.0279603)(1,0.0279603) 
};

\addplot [
color=red,
solid,
line width=1.2pt,
]
coordinates{
 (0.001,4.12982e-06)(0.0213878,0.00105391)(0.0417755,0.00281578)(0.0621633,0.00454607)(0.082551,0.00595814)(0.102939,0.00696256)(0.123327,0.00756698)(0.143714,0.00782782)(0.164102,0.00782281)(0.18449,0.00763515)(0.204878,0.00734417)(0.225265,0.00702019)(0.245653,0.006722)(0.266041,0.00649574)(0.286429,0.00637513)(0.306816,0.00638226)(0.327204,0.0065288)(0.347592,0.00681743)(0.36798,0.00724344)(0.388367,0.00779621)(0.408755,0.00846069)(0.429143,0.00921876)(0.449531,0.0100506)(0.469918,0.0109353)(0.490306,0.0118524)(0.510694,0.0127822)(0.531082,0.0137062)(0.551469,0.0146081)(0.571857,0.0154735)(0.592245,0.0162906)(0.612633,0.0170498)(0.63302,0.017744)(0.653408,0.0183686)(0.673796,0.0189212)(0.694184,0.0194015)(0.714571,0.0198109)(0.734959,0.0201529)(0.755347,0.0204317)(0.775735,0.0206531)(0.796122,0.0208236)(0.81651,0.0209502)(0.836898,0.0210401)(0.857286,0.0211006)(0.877673,0.0211386)(0.898061,0.0211603)(0.918449,0.0211711)(0.938837,0.0211755)(0.959224,0.0211766)(0.979612,0.0211767)(1,0.0211767) 
};

\addplot [
color=mycolor2,
solid,
line width=1.2pt,
]
coordinates{
 (0.001,4.12625e-06)(0.0213878,0.00103679)(0.0417755,0.00272039)(0.0621633,0.00429437)(0.082551,0.00547027)(0.102939,0.00616359)(0.123327,0.00639056)(0.143714,0.00621788)(0.164102,0.00573422)(0.18449,0.00503374)(0.204878,0.00420623)(0.225265,0.00333179)(0.245653,0.00247821)(0.266041,0.00169962)(0.286429,0.00103661)(0.306816,0.000517336)(0.327204,0.000158308)(0.347592,-3.37828e-05)(0.36798,-6.06573e-05)(0.388367,6.92521e-05)(0.408755,0.00034229)(0.429143,0.000741041)(0.449531,0.00124567)(0.469918,0.00183499)(0.490306,0.00248747)(0.510694,0.00318213)(0.531082,0.00389906)(0.551469,0.00462008)(0.571857,0.00532894)(0.592245,0.00601173)(0.612633,0.00665694)(0.63302,0.00725545)(0.653408,0.00780055)(0.673796,0.00828801)(0.694184,0.00871566)(0.714571,0.00908321)(0.734959,0.00939237)(0.755347,0.00964612)(0.775735,0.00984878)(0.796122,0.0100056)(0.81651,0.0101227)(0.836898,0.0102062)(0.857286,0.0102626)(0.877673,0.0102982)(0.898061,0.0103186)(0.918449,0.0103288)(0.938837,0.0103329)(0.959224,0.010334)(0.979612,0.010334)(1,0.010334) 
};

\addplot [
color=mycolor3,
solid,
line width=1.2pt,
]
coordinates{
 (0.001,4.12457e-06)(0.0213878,0.00102866)(0.0417755,0.00267507)(0.0621633,0.00417474)(0.082551,0.00523828)(0.102939,0.00578349)(0.123327,0.00583063)(0.143714,0.00545125)(0.164102,0.00473918)(0.18449,0.00379375)(0.204878,0.00270974)(0.225265,0.00157188)(0.245653,0.000452253)(0.266041,-0.000591227)(0.286429,-0.00151464)(0.306816,-0.00228693)(0.327204,-0.00288925)(0.347592,-0.00331295)(0.36798,-0.00355829)(0.388367,-0.00363262)(0.408755,-0.00354887)(0.429143,-0.00332409)(0.449531,-0.00297804)(0.469918,-0.0025321)(0.490306,-0.0020082)(0.510694,-0.00142787)(0.531082,-0.000811771)(0.551469,-0.000178902)(0.571857,0.000453596)(0.592245,0.00107086)(0.612633,0.00166039)(0.63302,0.0022121)(0.653408,0.00271833)(0.673796,0.0031739)(0.694184,0.00357577)(0.714571,0.00392279)(0.734959,0.0042159)(0.755347,0.00445736)(0.775735,0.00465084)(0.796122,0.00480103)(0.81651,0.0049134)(0.836898,0.00499374)(0.857286,0.00504816)(0.877673,0.00508251)(0.898061,0.00510226)(0.918449,0.00511218)(0.938837,0.00511619)(0.959224,0.00511722)(0.979612,0.00511731)(1,0.00511731) 
};

\draw[<->,very thick] (axis description cs:0.3,0.12) -- node[right=0.1pt]{\footnotesize{$\Delta E$}} (axis description cs:0.3,0.28);
\node[] at (axis description cs:0.91,0.78) {\footnotesize{$\h \! = \! \h^{\mathrm{BP}}$}};
\node[] at (axis description cs:0.90,0.43) {\footnotesize{$\h \! = \! \h^{\mathrm{MAP}}$}};

\end{axis}
\end{tikzpicture}


%% file: coupled_system.tex

\label{section:coupledsystem}

The potential theory from Section \ref{section:singlesystem} is now extended to spatially-coupled systems.
Vectors of measures are denoted by underlines (e.g., $\msbx$) with $[\msbx]_{i} = \msx_{i}$.
Functionals operating on single densities are distinguished from those operating on vectors by their input (i.e., $U(\msx;\msc)$ vs.~ $U(\msbx;\msc)$).
Also, for vectors $\msbx$ and $\msbx'$, we write $\msbx \degreq \msbx'$ if $\msx_i \degreq \msx_i'$ for all $i$, and $\msbx \degr \msbx'$ if $\msbx \degreq \msbx'$ and $\msx_i \degr \msx_i'$ for some $i$.

\subsection{Spatial Coupling}
\label{sec:spatial_coupling}

The ideas underlying spatial coupling now appear to be quite general.
The local coupling in the system allows the effect of the perfect information, provided at the boundary, to propagate throughout the system.
In the large system limit, these coupled systems show a significant performance improvement.

The $(\lambda,\rho,N,w)$ SC ensemble is defined as follows.
A collection of $2N$ variable-node groups are placed at all positions in $\mc{N}_v=\{ 1, 2, \ldots, 2N \}$ and a collection of $2N+(w-1)$ check-node groups are placed at all positions in $\mc{N}_c=\{ 1, 2, \ldots, 2N+(w-1) \}$.
For notational convenience, the rightmost check-node group index is denoted by $\chend \triangleq 2N+ (w-1)$.
The integer $M$ is chosen large enough so that i) $ML_{i}$, $ML'(1)R_{j}/R'(1)$ are natural numbers for $1 \leq i \leq \mathrm{deg}(L)$, $1 \leq j \leq \mathrm{deg}(R)$, and ii) $ML'(1)$ is divisible by $w$.

At each variable-node group, $ML_{i}$ nodes of degree $i$  are placed for $1\leq i \leq \mathrm{deg}(L)$.
Similarly, at each check-node group, $ML'(1)R_{i}/R'(1)$ nodes of degree $i$ are placed for $1\leq i \leq \mathrm{deg}(R)$.
At each variable-node and check-node group the $ML'(1)$ sockets are partitioned into $w$ equal-sized groups using a uniform random permutation.
Denote these partitions, respectively, by $\mathcal{P}^{v}_{i_1,j}$ and $\mathcal{P}^{c}_{i_2,j}$ at variable-node and check-node groups, where $1 \leq i_1 \leq 2N$, $1 \leq i_2 \leq \chend$ and $1 \leq j \leq w$.
The SC system is constructed by connecting the sockets from $\mathcal{P}^{v}_{i,j}$ to $\mathcal{P}^{c}_{i+j-1,j}$ using uniform random permutations.
This construction leaves some sockets of the check-node groups at the boundaries unconnected and these sockets are assigned the binary value 0 (i.e., the socket and edge are removed).
These 0 values form the seed that gets decoding started.

\vspace{-1mm}

\subsection{Spatially-Coupled Systems}

Let $\msxvn_{i}^{(\ell)}$ be the variable-node output distribution at node $i$ after $\ell$ iterations of message passing.
Then, the input distribution to the $i$-th check-node group is the normalized sum of averaged variable-node output distributions
\begin{equation}
  \label{coupling}
  \msx_{i}^{(\ell)} = \frac{1}{w} \sum_{k=0}^{w-1} \msxvn_{i-k}^{(\ell)}.
\end{equation}
Backward averaging follows naturally from the setup and is essentially the transpose of the forward averaging for the check-node output distributions.
This model uses uniform coupling over a fixed window, but in a more general setting window size and coefficient weights could vary from node to node.
By virtue of the boundary conditions, $\msxvn_{i}^{(\ell)} = \cnunit$ for $i \notin \mc{N}_v$ and all $\ell$, and from the relation in (\ref{coupling}), this implies $\msx_{i}^{(\ell)} = \cnunit$ for $i \notin \mc{N}_c$ and all $\ell$.

Generalizing~\cite[Eqn. 12]{Kudekar-arxiv12} to irregular codes gives evolution of the variable-node output distributions,
\begin{equation}
  \label{equation:DE_update_0}
  \msxvn_{i}^{(\ell+1)} = \msc \vnop \lambda^{\vnop} \left(\frac{1}{w} \sum_{j=0}^{w-1} \rho^{\cnop} \left( \frac{1}{w}  \sum_{k=0}^{w-1} \msxvn_{i+j-k}^{(\ell)} \right) \right).
\end{equation}

Making a change of variables, the variable-node output distribution evolution in (\ref{equation:DE_update_0}) can be rewritten in terms of check-node input distributions
\begin{equation}
  \label{equation:DE_update_1}
  \msx_{i}^{(\ell+1)} = \frac{1}{w} \sum_{k=0}^{w-1} \msc_{i-k} \vnop \lambda^{\vnop} \left(  \frac{1}{w} \sum_{j=0}^{w-1} \rho^{\cnop} \left( \ms{x}_{i-k+j}^{(\ell)} \right) \right) ,
\end{equation}
\noindent where $\msc_{i} = \msc$ when $i \in \mc{N}_v$ and $\msc_{i}=\cnunit$ otherwise. 
While (\ref{equation:DE_update_0}) is a more natural representation for the underlying system, (\ref{equation:DE_update_1}) is more mathematically tractable and easily yields a coupled potential functional.
As such, we adopt the system characterized by (\ref{equation:DE_update_1}) and refer to it as the \emph{two-sided spatially-coupled system}.

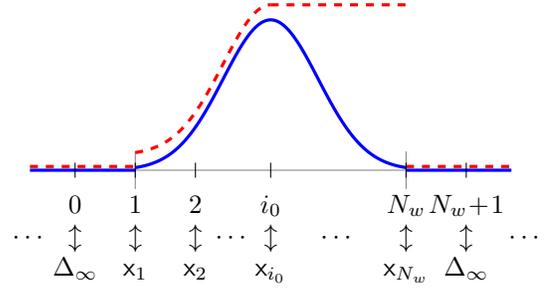
\begin{figure}[t]
  \centering
  \input{./Figures/SystemComparison}
  \vspace{-2mm}
  \caption{ \label{figure:system_comparison}
    This figure depicts the entropies of $\msx_1,\cdots,\msx_{\chend}$ in a typical iteration. 
    The blue line (solid) corresponds to the two-sided system and the red line (dashed) to the one-sided system.
    The distributions of the one-sided system are always degraded with respect to the two-sided system, hence a higher entropy.
    The distributions outside the set $\{1,\cdots,\chend\}$ are fixed to $\Delta_\infty$ for both the systems.
  } 
\end{figure}
\begin{figure*}[t]
  \begin{align}
    \label{equation:coupled_potential}
    U(\msbx ; \msc) \triangleq L'(1) \sum_{i=1}^{\chend}\left[ \frac{1}{R'(1)}  \ent{ R^{\cnop}( \ms{x}_{i}) } + \ent{ \rho^{\cnop}( \ms{x}_{i}) } - \ent{ \ms{x}_{i} \cnop \rho^{\cnop}(\msx_{i}) } \right] - \sum_{i=1}^{2N}   \ent{ \msc \vnop L^{\vnop} \Big{(} \frac{1}{w}  \sum_{j=0}^{w-1} \rho^{\cnop} ( \ms{x}_{i+j}) \Big{)} } 
  \end{align}
  \vspace{-4mm}
\end{figure*}
\begin{figure*}[t]
  \begin{align}
    \label{equation:first_derivative_coupled_system}
    \deri{\msbx} U(\msbx ; \msc)[\msby] =  L'(1) \sum_{i=1}^{\chend} \ent{ \left[ \msx_{i}  - \frac{1}{w} \sum_{k=0}^{w-1} \msc_{i-k} \vnop \lambda^{\vnop} \left( \frac{1}{w}  \sum_{j=0}^{w-1} \rho^{\cnop} (\msx_{i-k+j})   \right) \right] \vnop \left[ \rho'^{\cnop}(\msx_i) \cnop \msy_i \right] }
  \end{align}
  \vspace{-6mm}
\end{figure*}
\begin{figure*}[t]
  \begin{align}
    \label{equation:second_derivative_coupled_system}
    & \dderi{\msbx} U(\msbx ; \msc)[\msby,\msbz] = \notag \\  
    & L'(1) \rho''(1) \sum_{i=1}^{\chend} \ent{ \Bigg{[} \frac{1}{w} \sum_{k=0}^{w-1} \msc_{i-k} \vnop \lambda^{\vnop} \Bigg{(} \frac{1}{w} \sum_{j=0}^{w-1} \rho^{\cnop}( \msx_{i-k+j}) \Bigg{)} \cnop  \frac{\rho''^{\cnop}(\msx_{i})}{\rho''(1)} \Bigg{]} \cnop \msy_{i} \cnop \msz_{i} } \notag \\
    & - L'(1) \rho''(1) \sum_{i=1}^{\chend} \ent{ \left[ \msx_{i} \cnop \frac{\rho''^{\cnop}(\msx_{i})}{\rho''(1)}  \right] \cnop \msy_{i} \cnop \msz_{i} } - L'(1) \rho'(1) \sum_{i=1}^{\chend} \ent{ \frac{ \rho'^{\cnop}(\msx_{i}) }{\rho'(1)} \cnop \msy_{i} \cnop \msz_{i} }  \\
    & - \frac{L'(1)\lambda'(1) \rho'(1)^{2}}{w} \sum_{i=1}^{\chend} \sum_{ m = \max\{i-(w-1),1\} }^{ \min \{i+(w-1),\chend \}   } \mathrm{H} \Bigg{(} \frac{1}{w} \sum_{k=0}^{w-1}\msc_{i-k} \vnop \tfrac{ \lambda'^{\vnop} \left( \frac{1}{w} \sum\limits_{j=0}^{w-1} \rho^{\cnop}( \ms{x}_{i-k+j})\right)}{\lambda'(1)}  \notag \vnop \Big{[} \tfrac{\rho'^{\cnop}(\msx_{m})}{\rho'(1)}  \cnop \msz_{m} \Big{]}  \vnop \Big{[} \tfrac{\rho'^{\cnop}( \msx_{i})}{\rho'(1)}  \cnop \msy_{i} \Big{]} \Bigg{)} \notag
  \end{align}
  \vspace{-3mm}
  \noindent\makebox[\linewidth]{\rule{17.8cm}{0.4pt}}
\end{figure*}

The two-sided spatially-coupled system is initialized with
\begin{align*}
  \msx_{i}^{(0)} = \vnunit, \quad 1 \leq i \leq \chend .
\end{align*}
This symmetric initialization, uniform coupling coefficients, and symmetric boundary conditions (i.e., seed information) induce symmetry on all the message distributions.
In particular, the two-sided system is fully described by only half the distributions because
\begin{align*}
  \msx_{i}^{(\ell)} = \msx_{2N+w-i}^{(\ell)},
\end{align*}
for all $\ell$.
As density evolution progresses, the perfect boundary information from the left and right sides propagates inward.
This propagation induces a nondecreasing degradation ordering on positions $1, \ldots, \lceil \chend / 2 \rceil$ and a nonincreasing ordering on positions $\lceil (\chend+1) / 2 \rceil, \ldots, \chend$~\cite[Def. 44]{Kudekar-arxiv12}.

The nondecreasing ordering by degradation introduces a degraded maximum at $i_{0} \triangleq N + \lfloor \frac{w}{2} \rfloor$, and this maximum allows one to define a modified recursion that upper bounds the two-sided spatially-coupled system.
\begin{definition} [{\cite{Kudekar-arxiv12}}]
  The \emph{one-sided spatially-coupled system} is a modification of $(\ref{equation:DE_update_1})$ defined by fixing the values of positions outside $\mc{N}_c ' \triangleq \{1,2,\ldots,i_0\}$, where $i_{0}$ is defined as above.
  As before, the boundary is fixed to $\cnunit$, that is $\msx_{i}^{(\ell)} = \cnunit$ for $i \not \in \mc{N}_{c}$ and all $\ell$.
  More importantly, it also fixes the values $\msx_{i}^{(\ell)} = \msx_{i_0}^{(\ell)}$ for $ i_0 < i \leq \chend $ and all $\ell$.
\end{definition}

The density evolution update for the one-sided system is identical to (\ref{equation:DE_update_1}) for the first $i_{0}$ terms, $\{ \msx_{1}^{(\ell)}, \msx_{2}^{(\ell)}, \ldots, \msx_{i_{0}}^{(\ell)}  \}$.  But, for the remaining terms, it simply repeats the distribution $\msx_{i_{0}}^{(\ell)}$ and this implies $\msx_{i}^{(\ell)} = \msx_{i_{0}}^{(\ell)}$ for $i_{0} < i \le \chend$ at every step.
The one-sided and two-sided systems are illustrated in Fig.~\ref{figure:system_comparison}.

\begin{lemma}
  \label{lemma:FPOfOneSidedSystem}
  For the one-sided spatially-coupled system, the fixed point resulting from the density evolution satisfies 
  \begin{align*}
    \msx_{i} \degreq \msx_{i-1}, \quad 1 \leq i \leq \chend
  \end{align*}
\end{lemma}
\begin{IEEEproof}
  See \cite[Section IV-D]{Kudekar-arxiv12}.
\end{IEEEproof}

\subsection{Spatially-Coupled Potential}

\begin{definition}
  \label{def:coupled_potential}
  The potential functional for a spatially-coupled system $U : \mc{X}^{\chend} \times \mc{X} \rightarrow \mathbb{R}$ is given by (\ref{equation:coupled_potential}).
\end{definition}


\begin{lemma}
  \label{lemma:first_derivative_coupled_system}
  For a spatially-coupled system, the directional derivative of the potential functional with respect to $\msbx \in \probs^{\chend}$ evaluated in the direction $\msby \in \dpros^{\chend}$ is given by (\ref{equation:first_derivative_coupled_system}).
\end{lemma}

\begin{IEEEproof}
  The proof is similar to the single system case and is omitted for brevity.
\end{IEEEproof}

\begin{lemma}
  \label{lemma:second_derivative_coupled_system}
  For a spatially-coupled system, the second-order directional derivative of the potential functional with respect to $\msbx$ evaluated in the direction $[\msby,\msbz] \in \dpros^{\chend} \times \dpros^{\chend}$ is given by (\ref{equation:second_derivative_coupled_system}).
\end{lemma}

\begin{IEEEproof}
  We skip the proof for brevity.
\end{IEEEproof}


%% file: Figures/SystemComparison.tex
\begin{tikzpicture}[scale=0.2][domain=-20:20]
  \draw[very thin,color=gray] (-9,-1.25) -- (-9,1.25) ; 
  \draw[very thin,color=gray] (9,-1.25) -- (9,1.25) ; 
  \draw[very thin,color=gray] (-16,0) -- (16,0) ; 

  \draw[color=blue,smooth,samples=100,domain=-9:9,very thick] plot (\x,{10*exp(-\x * \x / 20)});
  \draw[color=blue,very thick] (9,0) -- (16,0);
  \draw[color=blue,very thick] (-16,0) -- (-9,0);

  \draw[color=red,smooth,samples=100,domain=-9:0,very thick,dashed] plot (\x,{1+10*exp(-\x * \x / 20)});
  \draw[color=red,smooth,samples=100,domain=0:9,very thick,dashed] plot (\x,{11});
  \draw[color=red,very thick,dashed] (9,0.25) -- (16,0.25);
  \draw[color=red,very thick,dashed] (-16,0.25) -- (-9,0.25);

  \node at (-16,0) [below=20pt] {$\cdots$};
  \draw[color=black] (-13,-0.5) -- (-13,0.5) node[below=7pt] {$\begin{array}{c} 0 \\ \updownarrow \\ \Delta_\infty \end{array}$} ; 
  \draw[color=black] (-9,-0.5) -- (-9,0.5) node[below=7pt] {$\begin{array}{c} 1 \\ \updownarrow \\ \mathsf{x}_1 \end{array}$};
  \draw[color=black] (-5,-0.5) -- (-5,0.5) node[below=7pt] {$\begin{array}{c} 2 \\ \updownarrow \\ \mathsf{x}_2 \end{array}$};
  \node at (-2.5,0) [below=20pt]  {$\cdots$};
  \draw[color=black] (0,-0.5) -- (0,0.5) node[below=7pt] {$\begin{array}{c} i_0 \\ \updownarrow \\ \mathsf{x}_{i_0}  \end{array}$}; 
  \node at (4.5,0) [below=20pt] {$\cdots$};
  \draw[color=black] (9,-0.5) -- (9,0.5) node[below=7pt] {$\begin{array}{c} \chend \\ \updownarrow \\ \mathsf{x}_{\chend} \end{array}$};
  \draw[color=black] (13,-0.5) -- (13,0.5) node[below=7pt] {$\begin{array}{c} \chend\!+\!1 \\ \updownarrow \\ \Delta_{\infty} \end{array}$ };
  \node at (17,0) [below=20pt] {$\cdots$};
\end{tikzpicture}


%% file: maintheorem.tex

We now prove threshold saturation for the spatially-coupled LDPC$(\lambda,\rho)$ ensemble.
Consider a spatially-coupled system with potential functional $U : \mc{X}^{\chend} \times \mc{X} \rightarrow \mathbb{R}$ as in Definition~\ref{def:coupled_potential}, and a parameterization $\phi : [0,1] \rightarrow \mathbb{R}$
\begin{equation*}
\phi(t) = U(\msbx + t(\msbx' - \msbx);\msc) ,
\end{equation*}
where $\msbx$ is a non-trivial fixed point of the one-sided SC system resulting from DE.
The path endpoint $\msbx'$ is chosen to impose a small perturbation on $\msbx$.
For all channels better than the potential threshold, $\msc \upgr \msc(\h^*)$, and any non-trivial fixed point, it can be shown that the one-sided SC potential strictly decreases along this perturbation.
However, since a fixed point is also a stationary point of the potential functional, all variations in the potential up to the second-order can be made arbitrarily small by choosing a large coupling parameter $w$.
Thus, one obtains a contradiction to the existence of a non-trivial fixed point from the second-order Taylor expansion of $\phi(t)$.

These ideas are formalized below.
A right shift is chosen for the perturbation.
This shift operator is defined in Definition \ref{definition:shift_operator} and Lemmas \ref{lemma:potential_shift_bound}, \ref{lemma:onesided_fixedpoint} discuss its effect on the potential functional at this shift and the directional derivative along the shift direction $[\mb{S}(\msbx) - \msbx]$.
Lemma \ref{lemma:second_derivative_bound} establishes an important bound on the second-order directional derivative of the potential.
Theorem \ref{theorem:threshold_saturation} proves threshold saturation.

\begin{definition}
  \label{definition:shift_operator}
  The shift operator $\mb{S} : \probs^{\chend}  \rightarrow \probs^{\chend}$ is defined pointwise by
  \begin{align*}
    [\mb{S}(\msbx)]_{1} &\triangleq \cnunit, & [\mb{S}(\msbx) ]_i &\triangleq \msx_{i-1}, \quad \text{$2 \leq i \leq \chend$}.
  \end{align*}
\end{definition}

\begin{lemma}[{\cite[Lemma 4]{Yedla-istc12}}]
  \label{lemma:potential_shift_bound}
  Let $\msbx \in \probs^{\chend}$ be such that $\msx_{i} = \msx_{i_0}$, for $i_0 \leq i \leq \chend$.
  Then, the change in the potential functional for a spatially-coupled system associated with the shift operator is bounded by
  \begin{align*}
    U(\mb{S}(\msbx) ; \msc) - U(\msbx ; \msc) \leq - U(\msx_{i_0} ; \msc).
  \end{align*}
\end{lemma}
\begin{IEEEproof}
  Due to boundary conditions, $\msx_{i}=\msx_{i_0}$ for $i_0 \leq i \leq \chend$, the only terms that contribute to $U(\mb{S}(\msbx) ; \msc) - U(\msbx ; \msc)$ are given in (\ref{equation:potential_shift_bound}).
  \begin{figure*}[t]
    \begin{align}
      \label{equation:potential_shift_bound}
      U(\mb{S}(\msbx) ; \msc) - U(\msbx ; \msc) & = - \frac{L'(1)}{R'(1)} \ent{ R^{\cnop}(\msx_{\chend}) } - L'(1) \ent{ \rho^{\cnop}(\msx_{\chend}) } +  L'(1) \ent{ \msx_{\chend} \cnop \rho^{\cnop}(\msx_{\chend}) } \\
      & \qquad \qquad + \ent{ \msc \vnop L^{\vnop} \Big{(} \frac{1}{w} \sum_{j=0}^{w-1} \rho^{\cnop}(\msx_{2N+j}) \Big{)} } - \ent{ \msc \vnop L^{\vnop} \Big{(} \frac{1}{w}  \sum_{j=0}^{w-1} \rho^{\cnop} ( \ms{x}_{j}) \Big{)} }, \, \text{where $\msx_0=\cnunit$} \notag
    \end{align}
    \vspace{-4mm}
    \noindent\makebox[\linewidth]{\rule{17.8cm}{0.4pt}}
  \end{figure*}
  As the last $w$ values, that is $\msx_{i}$ for $2N \leq i \leq \chend$, are degraded with respect to $\msx_{\chend}$ and $\msx_{\chend}=\msx_{i_0}$, we have the desired result.
\end{IEEEproof}

\begin{lemma}
  \label{lemma:onesided_fixedpoint}
  If $\msbx \degr \underline{\cnunit} \triangleq (\cnunit,\ldots,\cnunit)$ is a fixed point of the one-sided spatially-coupled system, then
  \begin{equation*}
    \deri{\msbx} U( \msbx ; \msc)[\mb{S}(\msbx) - \msbx] = 0,
  \end{equation*}
  and $\msx_{i_0} \notin \mc{V}(\msc)$ (i.e., $\msx_{i_0}$ not in basin of attraction of $\cnunit$).
\end{lemma}
\begin{IEEEproof}
  See Appendix.
\end{IEEEproof}

\begin{lemma}
  \label{lemma:second_derivative_bound}
  Let $\msbx_{1} \in \probs^{\chend}$ be a vector of symmetric probability measures, and let $\msbx_{2} \in \probs^{\chend}$ be a vector of coupled check-node inputs ordered by degradation, $[\msx_{2}]_i \degreq [\msx_{2}]_{i-1}$, generated by coupling a vector of variable-node outputs $\msbxcn_{2} \in \probs^{2N}$
  \begin{equation*}
    [\msx_{2}]_i  = \frac{1}{w} \sum_{k=0}^{w-1} [\tilde{\msx}_{2}]_{i-k} .
  \end{equation*}
  The second-order directional derivative of $U(\msbx_{1} ; \msc)$ with respect to $\msbx_{1}$ evaluated along 
  $[\mb{S}(\msbx_{2}) - \msbx_{2}, \mb{S}(\msbx_{2}) - \msbx_{2}]$
  can be absolutely bounded with 
  \begin{equation*}
    \abs{ \dderi{\msbx_{1}} U(\msbx_{1} ; \msc)[\mb{S}(\msbx_{2}) - \msbx_{2},\mb{S}(\msbx_{2}) - \msbx_{2}] } \leq \frac{K_{\lambda,\rho}}{w},
  \end{equation*}
  where the constant
  \begin{equation*}
    K_{\lambda,\rho} \triangleq \gamma L'(1) \left( 2\rho''(1) + \rho'(1) + 2\lambda'(1) \rho'(1)^{2} \right)
  \end{equation*}
  is independent of $N$ and $w$.
\end{lemma}

\begin{IEEEproof}
  See Appendix.
\end{IEEEproof}

\begin{theorem}
  \label{theorem:threshold_saturation}
  Consider a family of BMS channels $\msc(\h)$ that is ordered by degradation and parameterized by the entropy, \h.
  For a spatially-coupled LDPC $(\lambda,\rho)$ ensemble with a coupling window $w  > K_{\lambda,\rho}  / (2 \Delta E(\msc(\h)))$, and a channel $\msc(\h)$ with $\h < \h^*$, the only fixed point of density evolution is $\underline{\cnunit}$.
\end{theorem}

\begin{IEEEproof}
  Consider a one-sided spatially-coupled system.
  Fix $w>K_{\lambda,\rho}  / (2 \Delta E(\msc(\h)))$.
  Suppose $\msbx \degr \underline{\cnunit}$ is a fixed point of density evolution.
  Let $\msby = \mb{S}(\msbx) - \msbx$ and $\phi : [0,1] \rightarrow \mbb{R}$ be defined by
  \begin{equation*}
    \phi(t) = U(\msbx + t \msby ; \msc(\h)),
  \end{equation*}
  it is important to note that, for all $t \in [0,1]$, $\msbx + t \msby = (1-t) \msbx + t \mb{S}(\msbx) $ is a vector of probability measures.
  By linearity of the entropy functional and binary operators $\vnop$ and $\cnop$, $\phi$ is a polynomial in $t$, and thus infinitely differentiable over the entire unit interval.
  The second-order Taylor series expansion about $t=0$ evaluated at $t=1$ provides
  \begin{equation}
    \label{equation:phi_taylor}
    \phi(1) = \phi(0) + \phi'(0)(1-0) + \tfrac{1}{2} \phi''(t_0)(1-0)^{2},
  \end{equation}
  for some $t_0 \in [0,1]$.
  The first and second derivatives of $\phi$ are characterized by the first- and second-order directional derivatives of $U$:
  \begin{align*}
    \phi'(t) &= \lim_{\delta \rightarrow 0} \frac{ U( \msbx  + (t+\delta)\msby ; \msc(\h) ) - U(\msbx + t \msby ; \msc(\h)) }{\delta} \\
    &= \deri{\msbx +t \msby} U( \msbx + t \msby; \msc(\h))[\msby] ,
  \end{align*}
  and similarly,
  \begin{align*}
    \phi''(t) = \dderi{\msbx+t\msby} U(\msbx+t\msby ; \msc(\h))[\msby,\msby] .
  \end{align*}
  Substituting and rearranging terms in (\ref{equation:phi_taylor}) provides
  \allowdisplaybreaks{
    \begin{align*}
      & \tfrac{1}{2} \dderi{\msbx+t_0 \msby} U (\msbx+t_0 \msby ; \msc(\h) ) [\msby,\msby]  \\
      & \quad = U(\mb{S}(\msbx) ; \msc(\h)) - U(\msbx ; \msc(\h))- \deri{\msbx} U( \msbx ; \msc(\h))[\mb{S}(\msbx) - \msbx] \\
      & \quad = U(\mb{S}(\msbx) ; \msc(\h)) - U(\msbx ; \msc(\h)) \quad \text{(Lemma \ref{lemma:onesided_fixedpoint})}  \\
      & \quad \leq - U(\msx_{i_{0}} ; \msc(\h))   \quad \text{(Lemma \ref{lemma:potential_shift_bound})} \\
      & \quad \leq - \Delta E(\msc(\h)). \quad \text{(Lemma \ref{lemma:onesided_fixedpoint} and Definition \ref{definition:basin_of_attraction})}
  \end{align*}
  }
  Taking an absolute value and applying the second-order directional derivative bound from Lemma \ref{lemma:second_derivative_bound} gives
  \begin{equation*}
    \Delta E(\msc(\h)) \le \frac{K_{\lambda,\rho}}{2 w} \qquad \Longrightarrow \qquad w \le \frac{K_{\lambda,\rho}}{2 \Delta E(\msc(\h))},
  \end{equation*}
  a contradiction.
  Hence the only fixed point of the one-sided system is $\underline{\cnunit}$.
  The densities of the one-sided system are degraded with respect to the two-sided system, and therefore, the only fixed point of the two-sided system is also $\underline{\cnunit}$.
\end{IEEEproof}


%% file: appendix.tex

In the appendix, for a vector of measures $\msbx_i$, $\msx_{i,j}$ is used to denote $[\msbx_{i}]_{j} $.

\subsection{Proof of Lemma \ref{lemma:onesided_fixedpoint}}
Since $\msbx$ is a fixed point of the one-sided spatially coupled system,
\vspace{-2mm}
\begin{align*}
  \msx_{i} = \frac{1}{w} \sum_{k=0}^{w-1} \msc_{i-k} \vnop \lambda^{\vnop} \left( \frac{1}{w} \sum_{j=0}^{w-1} \rho^{\cnop} (\msx_{i-k+j})  \right)
\end{align*}
for $1 \leq i \leq i_0 $, and  $[\mb{S}(\msbx) - \msbx]_i = 0$ for $i_{0} < i \le \chend$.
The first result follows from applying these relations to the directional derivative, given in Lemma \ref{lemma:first_derivative_coupled_system}. 
Now, observe that 
\begin{align*}
  \msx_{i_0} & = \frac{1}{w} \sum_{k=0}^{w-1} \msc_{i_0-k} \vnop \lambda^{\vnop} \left( \frac{1}{w} \sum_{j=0}^{w-1} \rho^{\cnop} (\msx_{i_0-k+j})  \right) \\
  & \upgreq \frac{1}{w} \sum_{k=0}^{w-1} \msc_{i_0-k} \vnop \lambda^{\vnop} \left( \frac{1}{w} \sum_{j=0}^{w-1} \rho^{\cnop} (\msx_{i_0})  \right) \\
  & \upgreq \msc \vnop \lambda^{\vnop} \left( \rho^{\cnop} \left( \msx_{i_0}  \right)  \right) .
\end{align*}
Hence $\msx_{i_0} \notin \mc{V}(\msc)$.

\subsection{Proof of Lemma \ref{lemma:second_derivative_bound}}
Let $\msby = \mb{S}(\msbx_2) - \msbx_2$, with componentwise decomposition
\begin{align*}
  \msy_{j} & = [\mb{S}(\msbx_{2}) - \msbx_{2}]_{j}  = \msx_{2,j-1} - \msx_{2,j}  \\
  & = \frac{1}{w} \sum_{k=0}^{w-1} \tilde{\msx}_{2,j-1-k} - \frac{1}{w} \sum_{k=0}^{w-1} \tilde{\msx}_{2,j-k} = \frac{1}{w}(\tilde{\msx}_{2,j-w} - \tilde{\msx}_{2,j}),
\end{align*}
where $\tilde{\msx}_{2,j} = \cnunit$ for $j < 1 $.
Referencing Lemma \ref{lemma:second_derivative_coupled_system}, the first three terms of the second-order directional derivative are of the form 
\begin{align*}
  & \ent{ \msx_{3} \! \cnop \! \msy_{i} \! \cnop \! \msy_{i} } \!=\! \frac{1}{w} \ent{ \msx_{3} \! \cnop \! ( \tilde{\msx}_{2,i} \! - \! \tilde{\msx}_{2,i-w}) \! \cnop \! (\msx_{2,i} \! - \! \msx_{2,i-1})} ,
\end{align*}
by linearity.
From Corollary \ref{corollary:entropy_bound_3variables}, this term is absolutely bounded by
\vspace{-1.5mm}
\begin{align*}
  & \abs{ \ent{ \msx_{3} \cnop \msy_{i} \cnop \msy_{i} } } \leq \frac{ 1 }{w}  \ent{ \msx_{2,i} - \msx_{2,i-1}}.
\end{align*}
\vspace{-1.5mm}
The final term is of the form 
\begin{align*}
  & \abs{ \ent{ \msx_{4} \vnop \left[ \msx_{5} \cnop  \msy_{m} \right] \vnop \left[ \msx_{6} \cnop \msy_{i} \right] } }  \\
  &\quad = \abs{ \ent{ \left[ \msx_{4} \vnop \left( \msx_{5} \cnop  \msy_{m} \right) \right] \cnop \left[ \msx_{6} \cnop \msy_{i} \right] } } \quad \text{(Proposition \ref{proposition:duality_difference})}\\
  &\quad = \abs{ \ent{ \msx_{6} \cnop \left[ \msx_{4} \vnop \left( \msx_{5} \cnop  \msy_{m} \right) \right]   \cnop \msy_{i} } } \\
  &\quad = \frac{1}{w} \abs{ \ent{ \msx_{6} \cnop \left[ \msx_8 - \msx_7 \right] \cnop [\msx_{2,i} - \msx_{2,i-1}] } } \\
  &\quad \leq \frac{1}{w} \ent{\msx_{2,i} - \msx_{2,i-1}}. \quad \text{(Corollary \ref{corollary:entropy_bound_3variables})}
\end{align*}
By telescoping, one observes
\vspace{-2mm}
\begin{align*}
  \sum_{i=1}^{\chend} \ent{\msx_{2,i} - \msx_{2,i-1}} = \ent{ \msx_{2,\chend} - \cnunit } \leq 1 .
\end{align*}
Consolidating the above observations provides the desired upper bound.
